%% file: main.tex
\documentclass[sigplan]{acmart}

\AtBeginDocument{%
  \providecommand\BibTeX{{%
    \normalfont B\kern-0.5em{\scshape i\kern-0.25em b}\kern-0.8em\TeX}}}

\copyrightyear{2021}
\acmYear{2021}
\setcopyright{acmcopyright}\acmConference[ICCPS '21]{ACM/IEEE 12th International Conference on Cyber-Physical Systems (with CPS-IoT Week 2021)}{May 19--21, 2021}{Nashville, TN, USA}
\acmBooktitle{ACM/IEEE 12th International Conference on Cyber-Physical Systems (with CPS-IoT Week 2021) (ICCPS '21), May 19--21, 2021, Nashville, TN, USA}
\acmPrice{15.00}
\acmDOI{10.1145/3450267.3450546}
\acmISBN{978-1-4503-8353-0/21/05}

\input{macros}

\begin{document}

\title[Routing for Transportation in the Face of the COVID-19 Pandemic]{Incentivizing Routing Choices for Safe and Efficient Transportation in the Face of the COVID-19 Pandemic}
\author{Mark Beliaev}
\affiliation{%
  \institution{Electrical and Computer Engineering\\UC Santa Barbara}
}
\email{mbeliaev@ucsb.edu}

\author{Erdem B\i y\i k}
\affiliation{%
  \institution{Electrical Engineering\\Stanford University}
}
\email{ebiyik@stanford.edu}

\author{Daniel A. Lazar}
\affiliation{%
  \institution{Electrical and Computer Engineering\\UC Santa Barbara}
}
\email{dlazar@ucsb.edu}

\author{Woodrow Z. Wang}
\affiliation{%
 \institution{Computer Science\\Stanford University}
}
\email{wwang153@stanford.edu}

\author{Dorsa Sadigh}
\affiliation{%
  \institution{Computer Science\\Stanford University}
}
\email{dorsa@cs.stanford.edu}

\author{Ramtin Pedarsani}
\affiliation{%
  \institution{Electrical and Computer Engineering\\UC Santa Barbara}
}
\email{ramtin@ece.ucsb.edu}

\renewcommand{\shortauthors}{Beliaev et al.}

\begin{abstract}
The COVID-19 pandemic has severely affected many aspects of people's daily lives. While many countries are in a re-opening stage, some effects of the pandemic on people's behaviors are expected to last much longer, including how they choose between different transport options. Experts predict considerably delayed recovery of the public transport options, as people try to avoid crowded places. In turn, significant increases in traffic congestion are expected, since people are likely to prefer using their own vehicles or taxis as opposed to riskier and more crowded options such as the railway. In this paper, we propose to use financial incentives to set the tradeoff between risk of infection and congestion to achieve safe and efficient transportation networks. To this end, we formulate a network optimization problem to optimize taxi fares. For our framework to be useful in various cities and times of the day without much designer effort, we also propose a data-driven approach to learn human preferences about transport options, which is then used in our taxi fare optimization. Our user studies and simulation experiments show our framework is able to minimize congestion and risk of infection.
\end{abstract}



\begin{teaserfigure}
  \includegraphics[width=\textwidth]{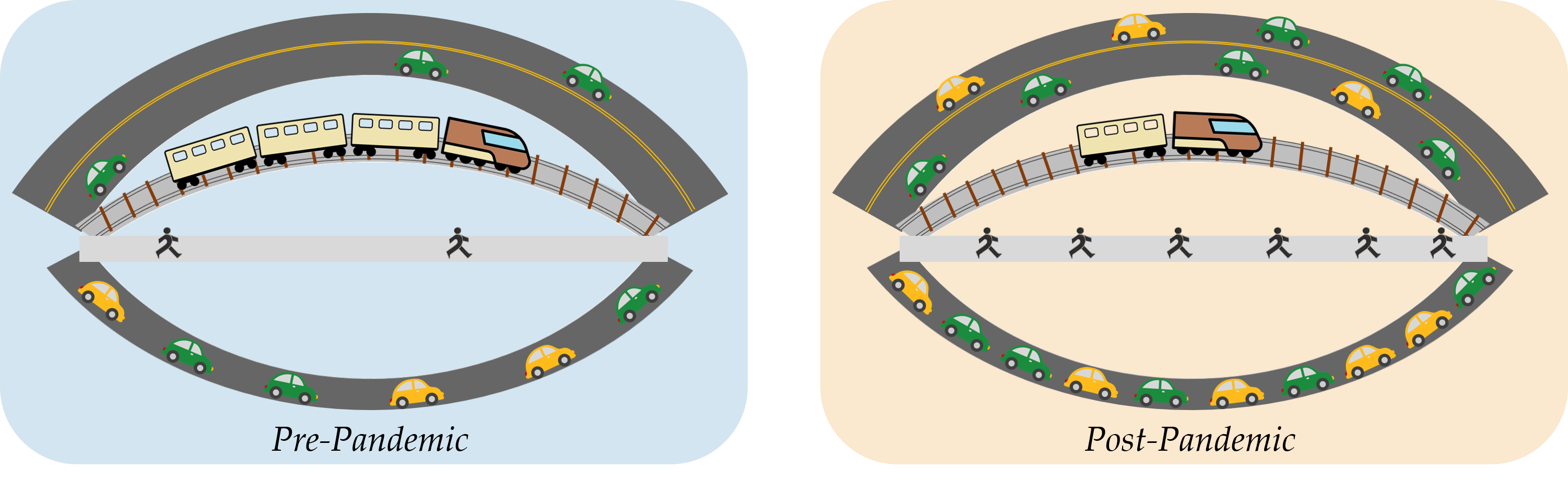}
  \vspace{-25px}
  \caption{The cartoons depict an example of a transportation network we investigate in this paper. People use four different modes of transportation: private cars (green), taxis (yellow), railway, and walking. On the right, people avoid using public transportation modes (railway in this case) due to the effects of pandemic. In addition to more people having to walk, more private cars and taxis operate in the network, which increases traffic congestion and travel delays for everyone.}
  \label{fig:front_fig}
  \vspace{5px}
\end{teaserfigure}

\maketitle
\section{Introduction}
As months go by, most of the world is still experiencing difficulties caused by the COVID-19 pandemic. With many governments pushing towards re-opening, experts predict more severe traffic congestion than before the pandemic~\cite{carpenter2020la}, especially in urban cities that rely much on public transportation~\cite{hu2020impacts}.
For example, studies suggest that in San Francisco, commuters will experience an additional round-trip delay of 20--80 minutes per person, at a societal cost of 556,000 -- 2,736,000 added traffic hours per day~\cite{hu2020the}.

One of the main reasons for expecting such a dramatic increase in traffic congestion is because the inhabitants of large metropolitan areas remain reluctant to use public transport~\cite{hertzke2020moving}, and transportation engineers predict a similar trend to continue for a long period of time~\cite{ewoldsen2020covid,pedersen2020impacts}. 
In fact, large-scale user surveys that aggregate responses from different countries show, for both business and private trips, \emph{risk of infection} has become the primary deciding factor on people's choice of the mode of transportation, becoming more important than time to destination, price of trip, avoidance of congestion, convenience, space, and privacy as compared to pre-pandemic conditions~\cite{hattrup2020five}.
Accordingly, the use of public transportation has substantially declined~\cite{tirachini2020covid}. While there has also been a drastic decrease in the use of private vehicles due to lockdowns, the decline has been much more drastic for public transportation~\cite{hertzke2020moving}, and mobility data show the recovery with re-opening takes much longer for public transportation~\cite{pascale2020here}. 
A similar shift away from ride-hailing services~\cite{vitale2020how} and taxis~\cite{zheng2020fall} is also expected, adding more challenges to post-pandemic traffic congestion.

While ``work from home" practices that have become more common during the pandemic help decrease the congestion for now~\cite{kahaner2020working}, many metropolitan cities have started to take other precautions to avoid significant increases in traffic congestion as well as to mitigate the risk of infection. For example, London and Paris have embarked on plans to create more biking and walking lanes and routes~\cite{hu2020will}, whereas Istanbul implemented a model of alternate working hours for public servants and schools~\cite{istanbul2020governor}. While different solutions may be applicable based on cultural factors or geographical conditions, researchers and experts emphasize financial incentives may play an important role for the recovery of public transportation and ride-hailing services~\cite{teale2020transportation}. To achieve this, we need to address the challenge of modeling how human preferences have evolved through the pandemic, and how these new preferences should affect routing and pricing public transportation and ride-hailing services.

Our insight in this paper is that \emph{we can leverage data-driven techniques to learn how humans' choices of mode of transportation have changed due to the pandemic, which we can use to optimize transportation networks in order to mitigate infection risk and decrease delays due to congestion}. By modeling the network with four modes of transportation, namely private cars, taxis (or vehicles of ride-hailing services), railway, and pedestrians, we optimize the taxi fares based on total demand and other network properties for a social objective consisting of two factors: safety and efficiency. The users of the network choose their mode of transportation and the routes selfishly, i.e., they choose the route that will minimize their own latency, monetary cost and risk of infection.

Our model can be seen as an indirect Stackelberg game~\cite{swamy2012effectiveness,krichene2017stackelberg}, where our planner first indirectly influences the demand for cars, taxis and railway by deciding the taxi fares. Then the humans respond selfishly by taking the highest-utility route available to them, where they might be optimizing based on their own priorities for latency, monetary cost, and risk of infection. Such a control scheme using financial incentives has previously been shown effective for mixed-autonomy traffic where regular and autonomous vehicles co-exist~\cite{biyik2019green}.

In order for our planning module to take into account people's preferences about the mode of transportation, we learn how people tradeoff latency, price and risk of infection using preference-based learning. We adopt active learning techniques to improve data-efficiency, as the human is in-the-loop. Our user study results demonstrate that people care more about the risk of infection while choosing their transport post-pandemic. In addition, our results on a simulated traffic network shows our network optimization can be utilized to set the tradeoff between congestion and the risk of infection.

Our contributions in this work are three-fold:
\begin{itemize}[nosep]
    \item We study different modes of transportation in traffic networks by modeling their latencies and risks of infection.
    \item We leverage active preference-based learning techniques to learn humans' preferences about the transport options while taking latency, monetary cost and the risk of infection into account.
    \item We formulate an optimization to set the taxi fares in the network, which uses the learned human preferences, to minimize traffic congestion and the risk of infection.
\end{itemize}


\section{Related Work}\label{sec:related_work}
We now overview the works that study modeling traffic under pandemic, routing games and pricing in transportation.

\smallskip
\noindent\textbf{Traffic under Pandemic.} Several works investigated the impact of the pandemic on transportation networks. \citet{wang2020spatial} analyzed the Multiscale Dynamic Human Mobility Flow Dataset \cite{kang2020multiscale} and Google Trends to investigate the correlation between the change of mobility patterns, government policies, and people's awareness in the United States.
\citet{cui2020traffic} proposed a traffic performance score to measure the impact of the pandemic on urban mobility. \citet{tirachini2020covid} pointed out the problems regarding public transportation due to the pandemic, and urged authorities for further research in order to sustain the benefits of public transportation. More related to our work, \citet{hu2020impacts} and \citet{zheng2020fall} showed the pandemic caused a change in the preferred transportation modes and the decline in the use of public transport may lead to worse traffic congestion than before.
While all of these works are important to understand the scale of the problem and motivate our work, they do not propose a solution for achieving safe and efficient mobility under the pandemic.

\smallskip
\noindent\textbf{Routing Games.} Previous works in traffic optimization have formulated the routing problem as a game between drivers. \citet{krichene2017stackelberg} developed an algorithm for efficiently finding \emph{Nash equilibria}, where drivers have no incentive to unilaterally deviate from their route choice, in parallel traffic networks under a single mode of transportation. Prior to that, \citet{roughgarden2002bad} and \citet{correa2008geometric} had formalized how bad the Nash equilibrium can be in a routing game. In this paper, we focus on a setting where owners of private cars reach a Nash equilibrium, whereas the routing of taxis are indirectly controlled by a social planner through pricing. This is similar to the \emph{altruistic} Nash equilibrium proposed by \citet{biyik2018altruistic}, except we do not make altruism assumptions, but use financial incentives to attain the benefits of altruism. \citet{lazar2019learning} used reinforcement learning to solve Stackelberg routing games with mixed-autonomy where the planner had full control over the autonomous vehicles. Most relatedly, \citet{biyik2019green} considered a similar setup in parallel networks where some portion of the traffic flow is indirectly controlled through pricing and the rest choose their routes selfishly. In this work, we extend the model to include public transportation (trains in particular) and incorporate safety measures due to the impacts of the pandemic, creating a more complex social objective.

\smallskip
\noindent\textbf{Pricing in Transportation Networks.} Our work is also related to the research in tolling \cite{beckmann1955studies}, as we use financial incentives to enhance safety and efficiency. Specifically, \citet{fleischer2004tolls} and \citet{brown2016robustness} studied tolling while taking the users' various preferences into account. \citet{sandholm2002evolutionary} derived tolls to have drivers choose socially optimal strategies. In the framework we consider in this work, train fares are constant and the pricing scheme is only for taxis that take various routes, rather than tolling.

\smallskip
\noindent\textbf{Learning Humans' Routing Preferences.} Finally, our work leverages preference-based learning which enable learning humans' preferences in the absence of user demonstrations \cite{sadigh2017active,kallus2016revealed,zadimoghaddam2012efficiently}. In preference-based learning, the user is queried with a set of options from which they are asked to select the option they most like. While \citet{biyik2019green} used preference-based learning with the volume removal objective to actively generate the queries, we adopt the information gain objective \cite{biyik2019asking}, which has been shown to be more data-efficient and user-friendly in terms of the easiness of queries.

\section{Transportation Network Model}\label{sec:model}
We begin by presenting the framework we use to model the traffic network we consider in this paper, as well as the monetary cost and risk of infection associated with each transportation mode. Our traffic network has one origin-destination (O-D) pair and multiple parallel roads that connect them. Roads are shared between private cars and taxis. Additionally, a railroad and pedestrian walking path connect the origin and destination.\footnote{Our framework easily generalizes to the networks with no railroad or no walking path. One will simply need to remove those options and the corresponding constraints.} This is visualized in Fig.~\ref{fig:front_fig}.

Our goal is to optimize the taxi fares, which depend on the route taken, to influence the state of the traffic to decrease latency and risk of infection.
While optimizing taxi fares, our framework will also estimate the latency and risk of infection associated with each transport option. These estimations are then revealed to the commuters so that they can take these factors into account while choosing their transport option.
In addition to the mode of transportation, people also choose their routes, i.e. they do not only choose to take a taxi, but also which route the taxi should take; because this has a direct effect on the fare they will pay. Such a solution can be feasible especially in ride-hailing services where customers choose their transport option prior to calling a vehicle.

There is an important tradeoff in our optimization: if taxi fares are too high, then more people may choose to take the train. While this decreases traffic congestion, high population density in the train may increase the risk of infection. On the other hand, too low taxi rates will hinder the recovery of public transport, increasing the traffic congestion. In addition to this tradeoff, allowing different fares for taxis based on the road they are taking gives more opportunities while still being fair to the customers\footnote{Fairness is satisfied because all taxi customers using the same road pay the same fare.}: they can be financially incentivized with lower fares to take longer roads, which may keep the congestion low by preventing the quickest roads from being overused.

To formulate our optimization for taxi fares, we elaborate on each transportation medium in the subsequent sections.

\subsection{Roads}
We consider $\numroads$ parallel roads and use $\setroads=\{1,2,\ldots,\numroads\}$ to denote the set of all roads. We denote the free-flow latency of road $i$, the time it takes to drive from origin to destination without any congestion,  by $\fflatency_i$.

\smallskip
\noindent\textbf{Latency.} Private cars and taxis share the roads, so they both contribute to road congestion. The latency of a road depends on how many vehicles are using it. We use the traffic model proposed by the Bureau of Public Roads (BPR model) \cite{united1964traffic,daskin1985urban} to mathematically formulate this latency. This model has been widely used in literature for traffic management \cite{dowling1998accuracy}, simulations \cite{florian1976recent}, urban-scale analysis \cite{wong2016network,kucharski2017estimating}, and modeling mixed-autonomy traffic \cite{mehr2019will,lazar2020routing,lazar2019optimal}.

We let $\flow_i^{c}$ denote the flow of private cars on road $i$, i.e., the number of private cars who use road $i$ per unit time. Similarly, $\flow_i^{t}$ denotes the flow of taxis on road $i$. Then the total flow of vehicles on road $i$ is $\flow_i^{v} = \flow_i^{c}+\flow_i^{t}$. According to the BPR model, the latency of this road is
\begin{equation}
\latency_i^{v}(\flow_i^{v}) = \fflatency_i\Big[1+\alpha\Big(\frac{\flow_i^{v}}{\capacity_i}\Big)^\beta\Big]\:,
\end{equation}
where $\fflatency_i$ and $\capacity_i$ are constants denoting the free-flow latency and capacity of the road, respectively, and $\alpha, \beta\in\mathbb{R}_{>0}$ are the parameters of the BPR model. The capacity $\capacity_i$ is proportional to the length and number of lanes of the road.

We then write the aggregate latency incurred per unit time on road $i$ and over the entire network as:
\begin{align}
    \Latency_i^{v}(\flow_i^{v}) = \flow_i^v \cdot \latency_i^{v}(\flow_i^{v})\:,\\ 
    \Latency^v(\flowvec^{v}) = \sum_{i\in\setroads}\Latency_i^{v}(\flow_i^{v})\:, 
\end{align}
respectively, where $\flowvec^v$ is the vector that consists of $\flow^v_i$ for $i=1,2,\dots,\numroads$. This is total latency incurred \emph{per unit time}, because $\flow_i^v$ denotes the \emph{flow}, i.e., number of people per unit time, using road $i$. 

\smallskip
\noindent\textbf{Monetary Cost.} Our goal is to optimize taxi fares for each road. We denote these fares as $\fare_i^t$ for $i\in\setroads$. On the other hand, we do not assume any control over the monetary cost of traveling with private cars, due to gas prices, tolling, etc. We neglect the deviations on the cost due to the effect of congestion on gas prices, and assume the cost is only affected by the characteristics of the road, e.g., its length and nominal vehicle speed. Therefore, $\fare_i^c$ is a constant for the monetary cost of taking road $i$ with a private car.

\smallskip
\noindent\textbf{Risk of Infection.} Private cars do not pose an extra threat in terms of infection, because they do not create new close interactions between people. On the other hand, taxis create an interaction between the passenger and the driver. Therefore, while the risk of infection is $0$ for private cars, we let $\bar\risk^t$ denote the risk of infection per unit time\footnote{Higher viral loads (the total number of virus particles taken in) might increase the severity of the disease. See, for example, \cite{westblade2020sars}.} for a single taxi. This value is affected by many factors that vary from city to city: the size and ventilation of taxis, whether there is a protective shield between the driver and the passenger, etc. And what is important is the ratio between $\bar\risk^t$ and the risk of infection in other transportation modes. Therefore, to keep our model general, we do not make additional assumptions about $\bar\risk^t$, which should be decided by health authorities.

Since longer interactions increase the risk, we model the total risk of infection a taxi causes during a trip on road $i$ as
\begin{equation}
    \risk_i^t(\flow_i^v) = \bar\risk^t \cdot \latency_i^v(\flow_i^v)\:,
\end{equation}
and total risk over the network per unit time due to taxis as
\begin{equation}
    \Risk^t(\flowvec^v) = \sum_{i\in\setroads}\flow_i^t \cdot \risk_i^t = \sum_{i\in\setroads}\flow_i^t \cdot \bar\risk^t \cdot \latency_i^v(\flow_i^v)\:.
\end{equation}
This is \emph{per unit time}, because $\flow_i^t$ denotes the \emph{flow}, i.e., number of people per unit time, using taxis. 

Having modeled the roads, we now continue with the railway in our transportation network.

\subsection{Railway}
Our network includes a railway from the origin to the destination for the train option. Railways have a physical capacity that determines at most how many people can take trains per unit time. We let this capacity be $\capacity^r$. We then define $\flow^r$ as the flow of people using the train, such that $0\leq\flow^r\leq\capacity^r$ must be satisfied.

\smallskip
\noindent\textbf{Latency.} Since trains operate on a schedule, the latency of a train is affected by neither how congested the roads are nor the number of people taking the train. Using railway from origin to destination takes a duration of constant $\latency^r$. Similar to roads, we can write the aggregate latency incurred per unit time on the railway as
\begin{equation}
    \Latency^r(\flow^r) = \flow^r \cdot \latency^r.
\end{equation}

\smallskip
\noindent\textbf{Monetary Cost.} Although the fares of taxis can be dynamically optimized based on the demand, railway has a fixed fare\footnote{Even though it may require important changes in the infrastructure, variable railway fares might be interesting to explore and provide additional benefits. Our formulation can be easily generalized to optimize for train fares in addition to taxi fares.}. We let $\fare^r$ denote the fare of the railway.

\smallskip
\noindent\textbf{Risk of Infection.} A very crucial aspect of the railway is that it may cause high risk of infection depending on how many people are aboard.
We let $\bar\risk^r$ denote the total risk of infection per unit time per person on a railway that operates at full capacity, i.e., when $\flow^r = \capacity^r$. Again letting health authorities decide $\bar\risk^r$ (relative to $\risk^t$), we can model the risk of infection per person  during one trip in the railway as
\begin{equation}
\risk^r(\flow^r)=\latency^r \cdot \bar\risk^r \cdot \frac{\flow^r}{\capacity^r} \:.
\end{equation}
The total risk of infection due to the railway per unit time is then written as
\begin{equation}
\Risk^r(\flow^r)=\flow^r \cdot \risk^r(\flow^r)=\latency^r \cdot \bar\risk^r \cdot \frac{{\flow^r}^2}{\capacity^r} \:.
\end{equation}

Next, we proceed to the model of the walking path to finalize our transportation network model.

\subsection{Walking Path}
Our network has a walking path from origin to destination for people who do not want to use the other modes of transport. We denote the flow of these people, pedestrians, as $\flow^p$. While we use a simple model for pedestrians, it again poses the same latency-risk tradeoff: having many people walk will reduce both the risk (compared to taxis and railway) and the congestion in the roads, but is not a desirable scenario, as pedestrians themselves will experience huge delays.

\smallskip
\noindent\textbf{Latency.} Dividing the path length by the average human walking speed, we assume the latency of a pedestrian is a constant, denoted by $\latency^p$. The aggregate latency incurred per unit time on the walking path is then
\begin{equation}
    \Latency^p(\flow^p) = \flow^p \cdot \latency^p \:.
\end{equation}

\smallskip
\noindent\textbf{Monetary Cost.} There is no monetary cost associated with walking to the destination.

\smallskip
\noindent\textbf{Risk of Infection.} The risk of infection for pedestrians depends on many external factors. For example, it increases if there are crowded places on the path, or it may decrease if policies enforce people to practice social distancing and wear face coverings. We denote the risk of infection per person per unit time on the path as $\bar\risk^p$, and so the total risk of infection per person on the walking path is $\risk^p = \latency^p\cdot \bar\risk^p$, and the total risk over the network per unit time is
\begin{equation}
    \Risk^p(\flow^p) = \flow^p \cdot \risk^p \:.
\end{equation}

Having modeled all the mediums in our network, we are now ready to formulate the optimization problem for taxi fares in order to influence routing decisions of people and bring down the total latency and risk of infection.

\section{Optimizing Safety \& Efficiency}\label{sec:optimization}
We first start with formulating our optimization objective, and then the constraints. Finally in Section~\ref{subsec:final_optimization}, we present the overall optimization problem.

\subsection{Objective}
Our goal is to optimize the taxi fares such that all people in the network will experience both low risk of infection and low latency. The network model we presented in Section~\ref{sec:model} allows us to model these two objectives:
\begin{align}
    \textrm{Latency: } & \Latency(\flowvec^v,\flow^r,\flow^p) = \Latency^v(\flowvec^v) + \Latency^r(\flow^r) + \Latency^p(\flow^p) \:, \label{eq:latency_objective}\\
    \textrm{Risk: } & \Risk(\flowvec^v,\flow^r,\flow^p) = \Risk^v(\flowvec^v) + \Risk^r(\flow^r) + \Risk^p(\flow^p) \:, \label{eq:risk_objective}
\end{align}
Our problem is then a multi-objective optimization where we try to minimize a weighted sum of \eqref{eq:latency_objective} and \eqref{eq:risk_objective}. Next, we formulate the constraints of our optimization.

\subsection{Constraints}
\noindent\textbf{Fare Constraints.} While we assume the city planner determines the taxi fares, we still need to make sure taxi drivers make profit, so we cannot set the fares arbitrarily low.

One can think of a solution where some minimum profit constraint is imposed over the entire network (similar to \cite{biyik2019green}). However, this requires some centralization to balance the profits because the optimization may produce very low fares for longer roads and high fares for shorter roads to satisfy the profit constraint.
In such a case, the taxis operating on the shorter roads would make a lot of profit while the other taxis are losing money. As taxis are usually decentralized, it is not realistic to assume a central authority will compensate the taxi drivers who lose money by taxing the ones who make more profit. This can, however, be an interesting direction for ride-hailing apps which naturally have a centralized system.

In this work, we instead impose the fare constraints separately for each road to make sure all taxis make profit. For this, we impose a minimum fare $\fare^t_i \geq \bar\fare^t_i$ for each road $i\in\setroads$ where $\bar\fare^t_i$ can be determined by how much minimum profit is desired for each trip on road $i$. One can think of $\bar\fare^t_i$ proportional to $\bar\fare^c_i$, depending on gas-efficiency of the taxis.

\smallskip
\noindent\textbf{Flow Constraints.} While we can financially incentivize people to take different modes of transportation by optimizing taxi fares, we cannot explicitly force them to take a particular mode. Hence, we are constrained by their preferences. To model humans' choices of the mode of transportation, we let $\prefratiovec$ be a discrete probability distribution over $2\numroads+2$ transport options: $n$ options with private cars, $n$ with taxis, $1$ with the railway and $1$ as a pedestrian. The vector $\prefratiovec$ sums up to one and consists of the following terms: $\prefratio^c_i$ for $i\in\setroads$, $\prefratio^t_i$ for $i\in\setroads$, $\prefratio^r$ and $\prefratio^p$. Each of these represents what fraction of the total flow chooses the corresponding transport option.

The vector $\prefratiovec$ of course depends on the latencies, monetary costs and the risks of infection of all transport options. Hence, we let it be a function: $\prefratiovec(\latencyvec,\farevec,\riskvec)$.
Here $\latencyvec$, $\farevec$ and $\riskvec$ denote the vectors that consist of the latency, monetary cost and risk of infection, respectively, for all transport options. For now, we treat the preference distribution $\prefratiovec(\latencyvec,\farevec,\riskvec)$ as a black-box function to complete our optimization formulation, and defer its derivation to Section~\ref{sec:preference_model}.

\subsection{Overall Optimization Problem}\label{subsec:final_optimization}
Letting $\demand$ denote the total demand per unit time, we can write the overall optimization problem as:
\begin{equation}
\begin{split}
    \min_{\farevec^t}\; & \gamma \cdot \Risk(\flowvec^v,\flow^r,\flow^p) + (1-\gamma) \cdot \Latency(\flowvec^v,\flow^r,\flow^p)\\
    \textrm{subject to }\; & \fare^t_i \geq \bar\fare^t_i \quad \forall i \in\setroads \:,\\
    & \latency_i^c = \latency_i^t = \fflatency_i\Big[1+\alpha\Big(\frac{\flow_i^c + \flow_i^t}{\capacity_i}\Big)^\beta\Big]\quad\forall i\in\setroads \:,\\
    & \risk^r = \latency^r \cdot \bar\risk^r \cdot \frac{\flow^r}{\capacity^r} \:,\quad \risk^p = \latency^p \cdot \bar\risk^p \:,\\
    &\risk_i^t = \bar\risk^t \cdot \latency_i^v \quad\forall i\in\setroads \:, \\
    & \flow_i^c = \demand \cdot \prefratio_i^c(\latencyvec,\farevec,\riskvec)\quad\forall i\in\setroads \:,\\
    & \flow_i^t = \demand \cdot \prefratio_i^t(\latencyvec,\farevec,\riskvec)\quad\forall i\in\setroads \:,\\
    & \flow^r = \demand \cdot \prefratio^r(\latencyvec,\farevec,\riskvec) \leq \capacity^r \:,\\
    & \flow^p = \demand \cdot \prefratio^p(\latencyvec,\farevec,\riskvec) \:,
\end{split}\label{eq:final_optimization}
\end{equation}
where $0\leq\gamma\leq1$ is a weight that determines the relative importance of the latency and risk objectives.

\section{Preference Distribution Model}\label{sec:preference_model}
The only missing component in our optimization is how we model the preference distribution $\prefratiovec(\latencyvec,\farevec,\riskvec)$. While all humans will presumably choose transport options to minimize their latency, monetary cost, and risk of infection, their tradeoff between these factors differ: some people prioritize minimizing their monetary cost whereas some prioritize reaching their destination as early as possible. Moreover, people may have other biases. For example, one may prefer taking a taxi rather than driving a car so that they can read during the trip, or one may prefer walking to stay healthy.

To incorporate such preferences, we model humans as agents optimizing a utility function that captures their preferences. For this, we first assume humans will not choose dominated options, i.e., options that have no advantage over some other option and have at least one disadvantage. Formally, option $j$, denoted by $\option_j$, is dominated by $\option_{j'}$ if:
\begin{gather}
    \option_j \textit{ and } \option_{j'} \textit{ are the same mode of transportation, }\textrm{and} \nonumber\\
    \left(\latency_{j'} \leq \latency_{j} \: \land \: \fare_{j'}\leq \fare_j \: \land \: \risk_{j'} \leq \risk_j\right), \textrm{ and} \\
    \left(\latency_{j'} < \latency_j \: \lor \: \fare_{j'} < \fare_j \: \lor \: \risk_{j'} < \risk_j\right) \:, \nonumber
\end{gather}
where $\latency_j$, $\fare_j$ and $\risk_j$ represent the latency, monetary cost and the risk of option $\option_j$, respectively, and similarly $\latency_{j'}$, $\fare_{j'}$ and $\risk_{j'}$ represent those of option $\option_{j'}$. Note that the railway and the walking path cannot be dominated as there is only one option each for them. The set of dominated options can be written as:
\begin{equation}
\begin{split}
D &= \{o_j \mid o_{j'} \textrm{ dominates } o_j \text{ for some option } \option_{j'} \}\:,
\end{split}
\end{equation}
Then, we model the utility of a human user, parameterized by $\weightvec\in\mathbb{R}^7$, as:
\begin{equation}
\begin{split}
    \utility_{\weightvec}(o_j) = \weight_1 &\ell_j + \weight_2 \fare_j + \weight_3 \risk_j + \\
    &\begin{cases}
    -\infty,\;\textrm{ if }o_j\in D,\\
    \weight_4,\;\textrm{ if }o_j\textrm{ is private car and }o_j\not\in D,\\
    \weight_5,\;\textrm{ if }o_j\textrm{ is taxi and }o_j\not\in D,\\
    \weight_6,\;\textrm{ if }o_j\textrm{ is railway},\\
    \weight_7,\;\textrm{ if }o_j\textrm{ is walking path}
    \end{cases} ,
\end{split}
\end{equation}
where $o_j$ denotes the chosen transport option. Here, the first three elements of $\weightvec$ characterize the tradeoff between the three factors of latency, cost, and risk, whereas the last four model humans' biases towards the different modes of transportation. Presumably, the first three elements of $\weightvec$ must be negative, penalizing high latency, cost and risk, for all humans. Such linear utility functions are common in preference-based learning as they are expressive enough to capture most of the important information \cite{wilde2020active,biyik2019asking,sadigh2017active}. Extensions to nonlinear models are possible with the use of Gaussian processes \cite{biyik2020active}.

Having defined a utility function for a human whose preferences are encoded with a vector $\weightvec$, we now adopt a widely-used probabilistic model from discrete choice theory to calculate the probability of the human choosing each option: multinomial logit model \cite{ben2018discrete}. This model has been extensively used both in transportation engineering \cite{koppelman1983predicting,biyik2019green} and preference-based learning \cite{sadigh2017active,biyik2019asking}. According to this model, the probability of the human choosing $\option_j$ among all options available to them, $\options$, is:
\begin{equation}
    P_{\weightvec}(\option_j\mid\options) = \frac{\exp(\utility_{\weightvec}(\option_j))}{\sum_{\option_{j'}\in\options} \exp(\utility_{\weightvec}(\option_{j'}))} \: .
\end{equation}
This model makes sure the dominated options have $0$ probability of being chosen, as their utilities are $-\infty$. Moreover, this model allows us to handle users with different transport options. For example, $\options$ may or may not include the private car option, depending on whether the user owns a car or not. We let $\options_k$ denote the options available to the user $k$.

We also let $\prefdist_k$ denote the distribution over user $k$'s preferences, i.e. their preference vector $\weightvec$. Then, the preference distribution $\prefratiovec(\latencyvec,\farevec,\riskvec)$ of the entire population of users can be written as:
\begin{equation}
    \prefratio_j(\latencyvec,\farevec,\riskvec) = \frac{1}{\population}\sum_{k=1}^{\population} \int_{\mathbb{R}^7} P_{\weightvec}(\option_j\mid\options_k)\prefdist_k(\weightvec)d\weightvec
\end{equation}
for each transport option $j$, where $\population$ is the population. This computation can be efficiently performed via sampling from $\prefdist_k$'s using, for example, Metropolis-Hastings algorithm.

In the next section, we discuss how data-driven methods can be leveraged to learn the distributions $\prefdist_k$, which will then complete our framework.

\section{Learning User Preferences}\label{sec:learning}
We take a Bayesian approach to learn the distributions $\prefdist_k$ of humans' preferences. In this approach, there exists a data set for each human that consists of their selected option as well as the options that were available to them while making their choice. 
While one can combine all data sets to learn a single $\prefdist$, learning a distribution for each human is advantageous, because this allows further personalizing our model to specific populations, such as people who travel in the early morning or later in the day. For example, people traveling early in the morning might care more about latency on their way to work, whereas travelers who leave later in the day may want to enjoy some nice weather by walking.
Therefore, we learn a separate distribution for each user, and $\prefdist$ is simply the average of such distributions associated with the humans traveling at that time.

Given a data sample for human $k$ where they chose option $\option_j$ among available options $\options_k$, we use Bayes' rule\footnote{The prior $\prefdist_k(\weightvec)$ depends on the city. While a simple prior is a uniform distribution over the $7$-dimensional unit ball, experts could incorporate their domain knowledge about the people using the network into this prior.} to update the posterior $\prefdist_k$:
\begin{equation}
    \prefdist_k\left(\weightvec \mid \option_j,\options_k\right) \propto \prefdist_k(\weightvec)P_{\weightvec}(\option_j \mid \options_k) \:.
\end{equation}
Assuming conditional independence of different choices, the posterior learned with the full data set of the human $k$ is:
\begin{equation}
    \prefdist_k\left(\weightvec \mid \dataset\right) \propto \prefdist_k(\weightvec)\prod_{(\option_j \mid \options_k)\in\dataset}P_{\weightvec}(\option_j \mid \options_k) \:,
\end{equation}
where $\dataset$ denotes the data set.

This learning approach yields better estimates of $\prefdist_k$ as we collect more data of human $k$. While we can generate more data by running our optimization in \eqref{eq:final_optimization} on real network instances, this optimization is only designed to optimize for taxi fares and is not designed for optimally collecting data from humans for learning their preferences. Hence, it does not necessarily generate queries (transport options) that will lead to significant improvements in our estimates of $\prefdist_k$, which may, in turn, hurt the performance of network optimization. To overcome this problem, we present an optional active querying method based on information gain in the subsequent section, which optimizes the queries, i.e. their latencies, monetary costs and risks without necessarily satisfying the network dynamics to learn the humans' preferences optimally. Such an approach could be employed in practice by, for example, having a survey for humans who enter the network for the first time.

\subsection{Active Querying} \label{subsec:active_learning}
To enable faster learning of $\prefdist_k$, we use an active learning method that optimizes the information gained from each query about the distribution $\prefdist_k$ of $\weightvec$. While prior works used a \emph{volume removal} based active learning approach where the goal is to maximize the volume removed from the distribution $g_k$ \cite{sadigh2017active,biyik2019green,golovin2011adaptive}, \citet{biyik2019asking} showed maximizing mutual information leads to both faster learning and easier queries for the humans.  In this method, each query is optimized after receiving the human's choice:
\begin{equation}
    \max_{\options_k} I(\weightvec ; \option_j \mid \options_k) \!=\! \max_{\options_k} \!\left[H(\weightvec \mid \options_k) \!-\! \mathbb{E}_{\option_j}\!\left[H(\weightvec \mid \option_j, \options_k)\right]\right],
    \label{eq:info_gain_optimization}
\end{equation}
where $I$ is the mutual information, $H$ is the information entropy \cite{cover1999elements}, and $\option_j$ is the random variable representing the selected transport option. After sampling $\weightvec\sim\prefdist_k(\weightvec)$ and letting the set of samples be $\Omega$, the optimization \eqref{eq:info_gain_optimization} is asymptotically equal to
\begin{equation}
    \max_{\options_k} \sum_{\option_j\in\options_k}\sum_{\weightvec\in\Omega}P_{\weightvec}(\option_j\mid\options_k)\log\frac{P_{\weightvec}(\option_j \mid \options_k)}{\sum_{\weightvec'\in\Omega}P_{\weightvec'}(\option_j \mid \options_k)}
\end{equation}
as the number of $\weightvec$ samples in $\Omega$ goes to infinity. We refer to \cite{biyik2019asking} for the full derivation.

We want to note again that this optimization is not subject to the constraints of the optimization we presented in \eqref{eq:final_optimization}, because the goal here is to learn the user preferences as quickly as possible using some artificial queries that ask the users about their transportation preferences.

Having presented our approach to learn humans' preferences over transport options and our active querying approach to accelerate learning, our framework is complete.

\section{Experiments \& Results} \label{sec:experiments}
In this section, we present our experiments and the results. We divide the experiments into two parts: a user study where we learn user preferences with the approach described in Section~\ref{sec:learning}, and a case study on a simulated traffic network where we optimize the taxi fares as formulated in Section~\ref{sec:optimization} using the preferences learned in the user study. 

\subsection{Learning User Preferences}\label{subsec:user_studies}
To perform the optimization for a given traffic network, we first need to learn the distribution of humans' preferences, $\prefdist_k$'s. For this, we surveyed 17 people across the United States during the second half of 2020, using the active learning framework we described in Section~\ref{subsec:active_learning}.

The participants were first explained the purpose of the survey, as well as the underlying model for calculating the risk of infection. Their place of residence, whether they were tested positive before for COVID-19, and their informed consent were collected.  Each participant then answered $10$ actively generated queries, followed by $6$ queries that were randomly generated. We used the latter set of queries to test the accuracy of the predicted utility parameters $\weightvec$: we looked at for how many of those queries our estimated $\weightvec$'s could correctly predict the actual participant response. Each query consisted of $6$ choices ---two roads with private cars and taxis, a railway, and a walking path. For the transport options, we provided the users with the estimated latencies, monetary costs and the estimated densities to give them an easily interpretable measure of infection risk. 

To validate the impact of the pandemic on humans' transport choices, we attempted to model the pre-pandemic distribution of people's preferences, as well. For this, we asked the same participants to repeat the survey as if they were making their decisions prior to the pandemic. All 17 participants attended this second round survey.

The validation accuracy of our learning model is $97.1\%$ for pre-pandemic conditions, and $89.8\%$ for post-pandemic. As each query consisted of 6 transport options, these high accuracy values indicate we were able to accurately model participants' preferences. The decrease in the accuracy from pre-pandemic to post-pandemic might be because risk of infection is less relevant in the pre-pandemic case,\footnote{Risk of infection is not completely irrelevant in the pre-pandemic case, as people would generally prefer transport options with fewer people due to comfort, hygiene, etc.} making the learning easier.

\begin{figure*}[ht]
    \centering
    \includegraphics[width=\textwidth]{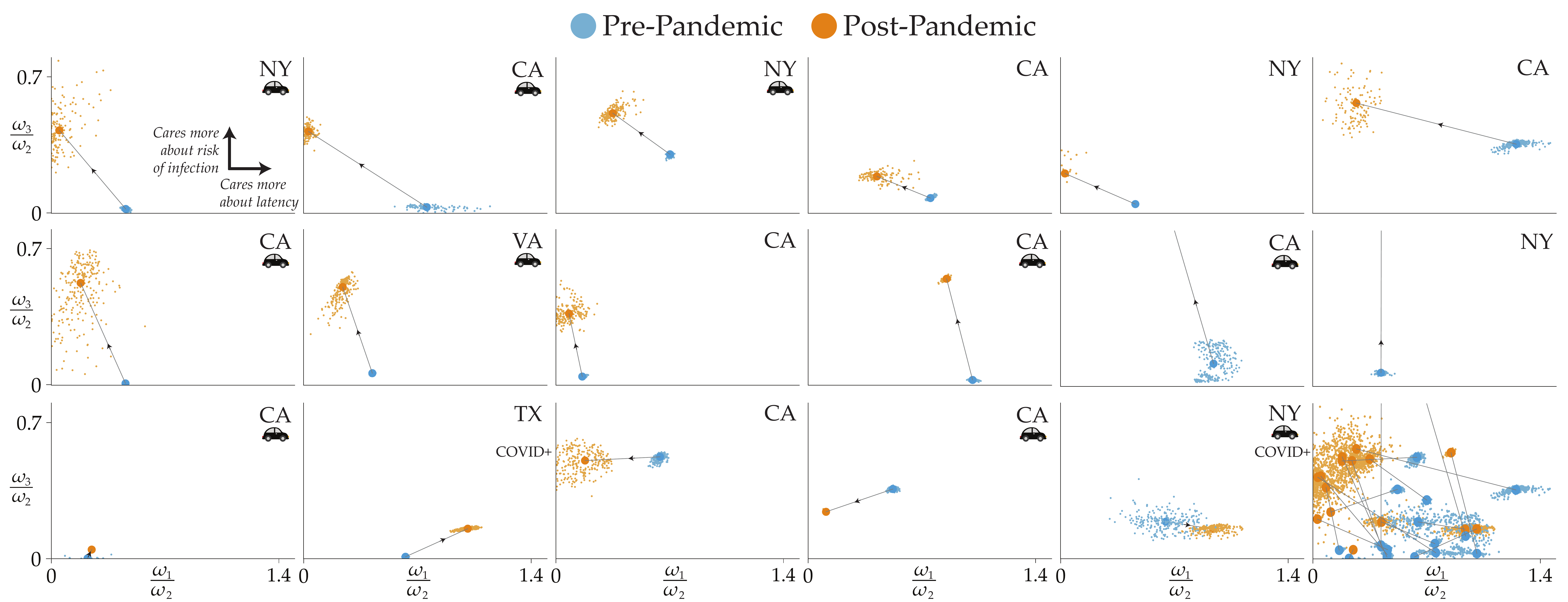}
    \caption{Distribution of participants' learned tradeoff parameters. Their places of residence are noted on the top right of each plot (CA - California, NY - New York, VA - Virginia, TX - Texas). Two participants who were infected with COVID-19 before have also been labeled. Car icons indicate whether those participants were simulated as private car owners or not in Section~\ref{subsec:simulated_results}. The bottom right plot shows all participants. It can be seen that $14$ of the $17$ participants care more about risk of infection in the post-pandemic case. The changes in the other three are much smaller.}
    \label{fig:parameters}
\end{figure*}

Using the preference data we collected, we generated two sets of preference distributions. Fig.~\ref{fig:parameters} visualize the tradeoffs for each of the 17 study participants, where small points show the samples from their $\prefdist_k(\weightvec)$, and the mean of the distribution is indicated with a large point.

We see from Fig.~\ref{fig:parameters} how the pandemic affected humans' relative tradeoffs between latency, monetary cost, and risk of infection. The majority of participants (14 out of 17) showed an increase in their regard for risk relative to monetary cost, and the changes in the other three are much smaller. This shows that people are putting a higher weight on the risk of infection associated with a travel option.

Moreover, the majority (13 out of 17) showed a decrease in their regard for latency relative to monetary cost. In the post-pandemic case, people are more willing to sacrifice their time for monetary considerations. This shows that our idea of using financial incentives to influence the state of the traffic network can in fact be effective.

It is interesting to point out that the only two participants whose preference ratios significantly moved rightward, i.e. who started caring more about latency, were the only two participants who were infected with and cured for COVID-19 prior to the surveys. This might be because they have less fear of being re-infected.

Having validated people are more concerned about the risk of infection, and shown the usability of our preference-based learning framework, we now proceed to the simulated traffic network experiments.

\subsection{Case Study} \label{subsec:simulated_results}
To show the benefits of our optimization framework, we designed a case study with a traffic network consisting of four modes of transportation. The network we used is similar to the depiction given in Fig.~\ref{fig:front_fig}, containing $\numroads=2$ roads, a railway, and a walking path, all connecting one O-D pair.

The free-flow latencies of the two roads are $\fflatency_1=30$ minutes and $\fflatency_2=45$ minutes. Their capacities are $\capacity_1=900$ and $\capacity_2=600$ vehicles per minute. The monetary cost associated with driving a private car on these roads is $\fare_1^c=\$15$ and $\fare_2^c=\$9$. The minimum taxi fares for the two roads are $\bar\fare_1^t=\$9$ and $\bar\fare_2^t=\$5$. Road 1 can be thought of as a high-capacity, high-cost freeway option, while road 2 is a low-capacity, low-cost street. The train latency, capacity, and fare are $\latency^r=35$ minutes, $\capacity^r=1500$ passengers per minute, and $\fare^r=\$3$ respectively. The walking path latency is $\latency^p=120$ minutes. 

We set the parameters related to risk of infection in our simulated problem as follows. Risks of infection per minute for taxis and pedestrians are $\bar\risk^t=\bar\risk^p=1$. For the railway, we set the risk of a full capacity train (per minute) as $\bar\risk^r=10$.

The flow demand for this network is set to $\demand=3000$ people per minute. We simulated this population using the $\weightvec$ samples we obtained from the user studies presented in Section~\ref{subsec:user_studies}. To separate the population of car-owners from the rest, as they will have different routing options $\options_k$, we simulated $10$ of the participants as car-owners. Those users are indicated with a car icon in Fig.~\ref{fig:parameters}.

\begin{figure}[ht]
    \centering
    \includegraphics[width=\columnwidth]{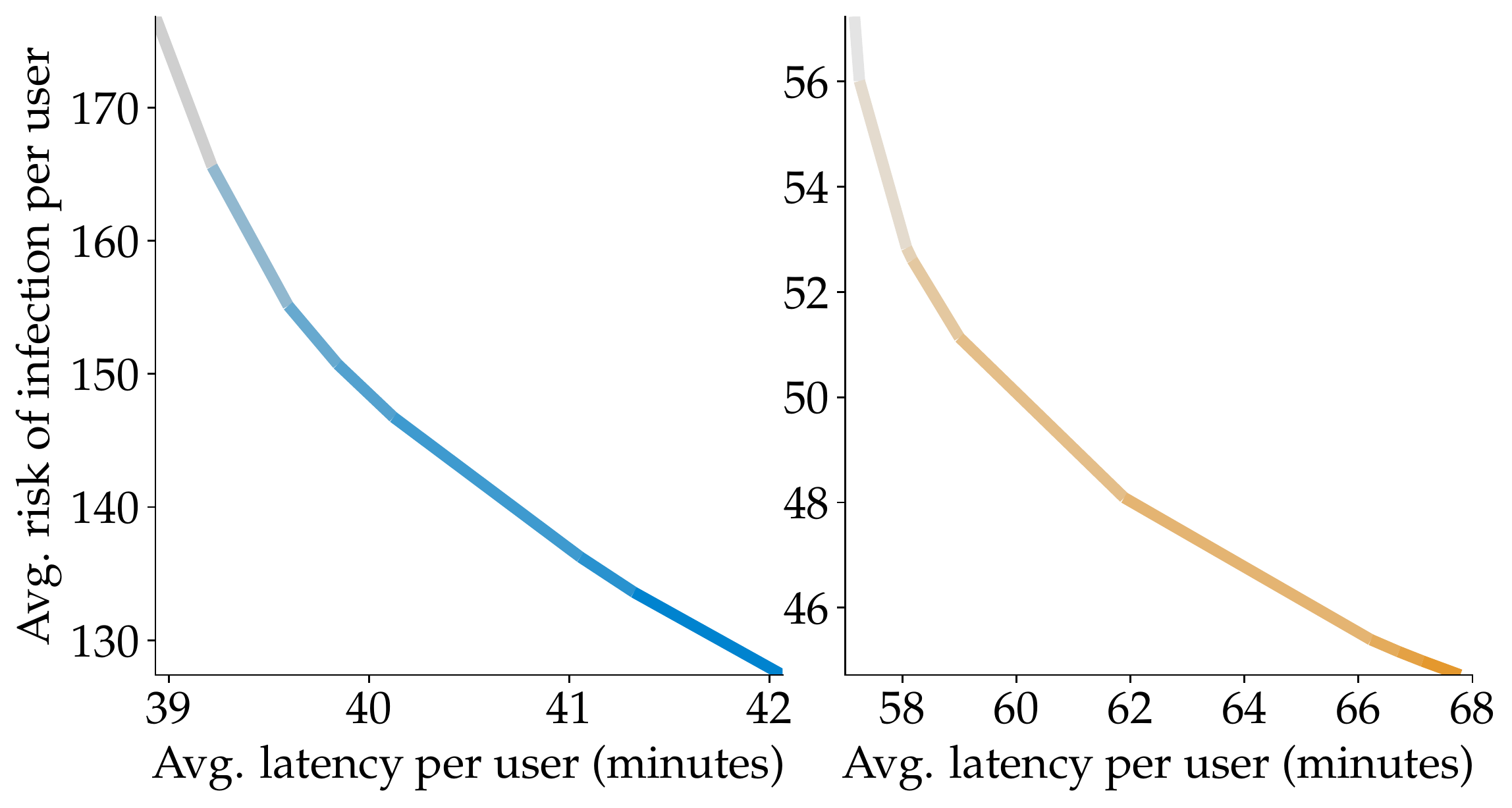}
    \caption{Optimized risk and latency tradeoff is shown for varying $\gamma$ under the pre-pandemic (left) and post-pandemic (right) conditions. Each value of $\gamma$ is used to generate a point on the plot, and line segments connecting adjacent points are colored to correspond with the values of $\gamma$. Darker colors correspond to higher $\gamma$. Higher $\gamma$ prioritizes minimizing risk of infection, at the cost of higher latencies.
    }
    \label{fig:portraits}
\end{figure}

We used a Sequential Quadratic Programming algorithm developed by \cite{slsqp} to locally solve the optimization, and repeated with 100 random initializations to get closer to the global optimum. We performed this procedure for the two data sets (pre- and post-pandemic) as well as varying values of $\gamma$. Fig.~\ref{fig:portraits} depicts the effect $\gamma$ has on the tradeoff between latency and risk of infection in our objective function. As expected, increasing $\gamma$ helps minimize risk of infection. The constraints enforced by users' post-pandemic preferences put higher priority on risk of infection as opposed to latency, making aggregate risk of infection less sensitive to changes in $\gamma$. On the other hand, for pre-pandemic preferences, people's disregard for risk and larger regard for latency make $\gamma$ have a larger effect on risk of infection as opposed to latency.

\begin{figure*}
    \includegraphics[width=\textwidth]{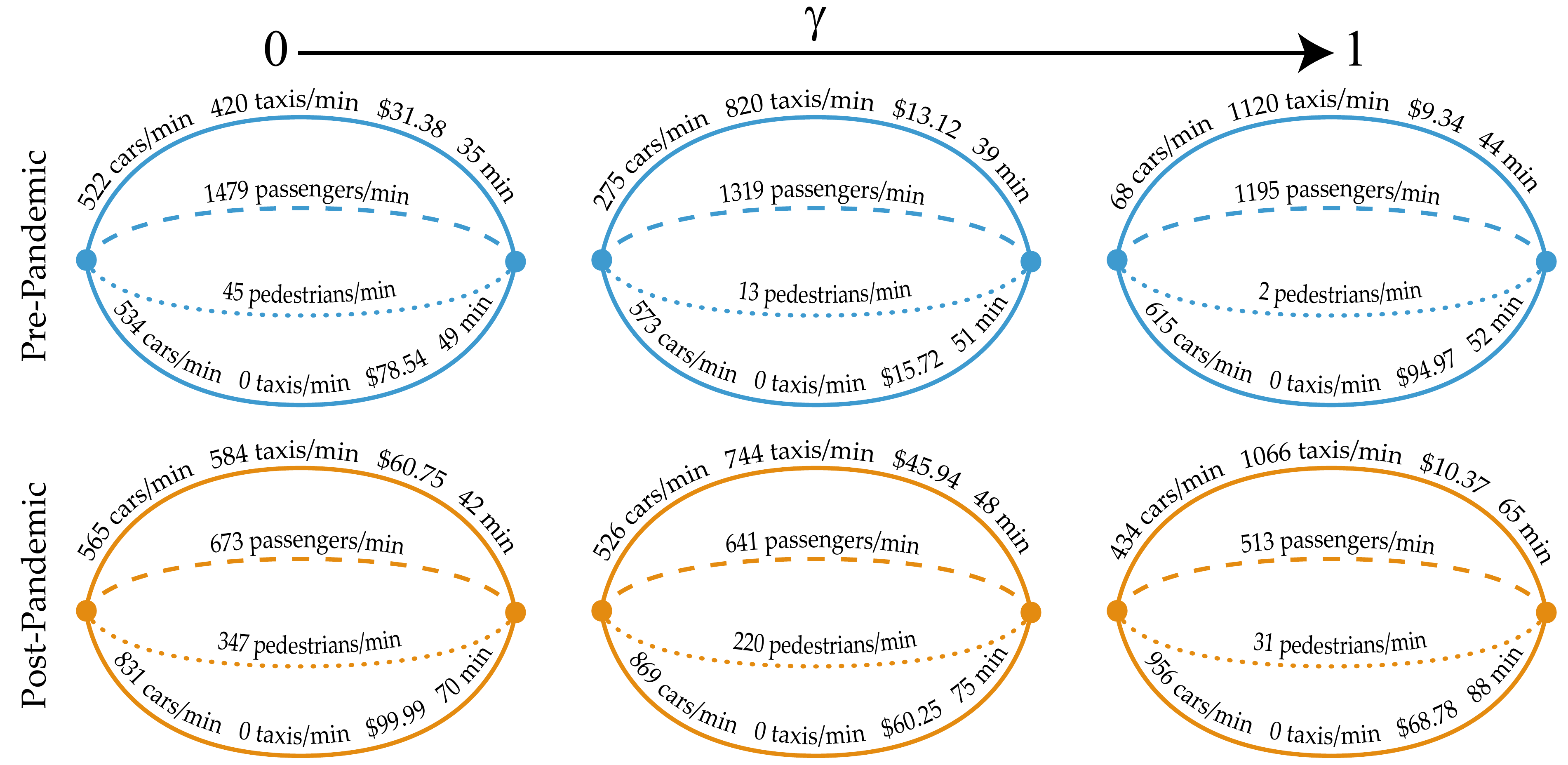}
    \caption{
      Solutions to the optimization for pre-pandemic (top) and post-pandemic (bottom) for varying $\gamma$ (increasing from left to right). The lines depict road 1, the railway, the walking path, and road 2 from top to bottom. As $\gamma$ increases, our framework fiscally incentivizes users to take the taxi on road $1$. This helps minimize the number of passengers on the railway, as well as the number of pedestrians on the walking path. Note that the price for a taxi on road $2$ does not have to follow a trend as $\gamma$ increases, but only has to stay higher than the price for a taxi on road $1$ so that it remains as a dominated option to make sure all the taxis go on road $1$.
    }
    \label{fig:solutions}
\end{figure*}

Finally, Fig.~\ref{fig:solutions} shows the specific solutions for three different $\gamma$ under pre-pandemic and post-pandemic conditions. Higher $\gamma$ leads to lower risk of infection by populating the railway less, whereas lower $\gamma$ decreases traffic congestion.

\section{Conclusion} \label{sec:conclusion}
\noindent\textbf{Summary.} We proposed a complete framework consisting of a learning and a planning module: we first learn humans' transport preferences in a data-driven way using active preference-based learning. We then use these learned preferences to optimize taxi fares in traffic networks. Our user study results align with the expert predictions \cite{hu2020will} and larger-scale surveys \cite{hattrup2020five}, and simulation experiments demonstrate the usability and the benefits of the framework.
\noindent\textbf{Discussion.} Due to the sensitive nature of the topic at hand, one should consider the moral implications that our results make. On the positive side, our results show that financial incentives can be used to decrease the risk of infection and accelerate the recovery of public transport. However, there is a natural tradeoff between these two objectives, which means hypothetically financial incentives could be used in ways that put riders at risk in order to recover public transport more quickly. Our goal here is not to advocate for such an approach, but to better understand people's preferences. We hope that authorities take the necessary precautions to decrease infection risks under a certain level, and use financial incentives only afterwards to recover public transport.

\noindent\textbf{Limitations and Future Work.} In this paper, we studied traffic networks that consist of parallel roads and a single O-D pair. While this is a common assumption (as in \cite{biyik2018altruistic,krichene2017stackelberg,biyik2019green}), extensions to general network topologies remain as a future work: one can think of enumerating all possible paths from all origins to all destinations and then calculating the path latencies as the sum of corresponding roads' latencies. In our case study, we considered a network with only two parallel roads for the clarity of presentation. While the real-world applications do not usually involve many parallel roads, our framework is scalable to the larger number of roads. Moreover, our user studies are limited due to the number of participants and the inherent bias people might have due to the pandemic while attending the pre-pandemic survey. Finally, we assumed the risk of infection is proportional to the interaction time in trains and taxis and for pedestrians. While this may not be accurate, our framework can be easily modified for better risk modeling.

\section*{Acknowledgments}
We acknowledge funding by NSF ECCS grant \#1952920.
 
\bibliographystyle{ACM-Reference-Format}
\balance
\bibliography{references}
\end{document}

%% file: macros.tex
\usepackage{bm}
\usepackage[font=small]{caption}
\usepackage{enumitem}
\setlist[itemize]{leftmargin=5.5mm}

\usepackage{titlesec}
\titlespacing*{\subsection}{0pt}{0.15\baselineskip}{0.05\baselineskip}
\titlespacing*{\section}{0pt}{0.6\baselineskip}{0.45\baselineskip}

\newcommand{\numroads}{n}
\newcommand{\setroads}{[\numroads]}

\newcommand{\fflatency}{a}
\newcommand{\flow}{f}
\newcommand{\flowvec}{\bm{\flow}}
\newcommand{\capacity}{c}
\newcommand{\latency}{\ell}
\newcommand{\latencyvec}{\bm{\latency}}
\newcommand{\Latency}{L}
\newcommand{\fare}{x}
\newcommand{\farevec}{\bm{\fare}}
\newcommand{\risk}{r}
\newcommand{\riskvec}{\bm{\risk}}
\newcommand{\Risk}{R}

\newcommand{\prefratio}{h}
\newcommand{\prefratiovec}{\bm{h}}

\newcommand{\demand}{F}

\newcommand{\weight}{\omega}
\newcommand{\weightvec}{\bm{\omega}}
\newcommand{\utility}{u}
\newcommand{\option}{o}
\newcommand{\options}{\mathcal{O}}
\newcommand{\prefdist}{g}
\newcommand{\population}{N}

\newcommand{\dataset}{\mathcal{D}}

%% file: main.bbl

\begin{thebibliography}{50}


\ifx \showCODEN    \undefined \def \showCODEN     #1{\unskip}     \fi
\ifx \showDOI      \undefined \def \showDOI       #1{#1}\fi
\ifx \showISBNx    \undefined \def \showISBNx     #1{\unskip}     \fi
\ifx \showISBNxiii \undefined \def \showISBNxiii  #1{\unskip}     \fi
\ifx \showISSN     \undefined \def \showISSN      #1{\unskip}     \fi
\ifx \showLCCN     \undefined \def \showLCCN      #1{\unskip}     \fi
\ifx \shownote     \undefined \def \shownote      #1{#1}          \fi
\ifx \showarticletitle \undefined \def \showarticletitle #1{#1}   \fi
\ifx \showURL      \undefined \def \showURL       {\relax}        \fi
\providecommand\bibfield[2]{#2}
\providecommand\bibinfo[2]{#2}
\providecommand\natexlab[1]{#1}
\providecommand\showeprint[2][]{arXiv:#2}

\bibitem[\protect\citeauthoryear{Beckmann, McGuire, and Winsten}{Beckmann
  et~al\mbox{.}}{1955}]%
        {beckmann1955studies}
\bibfield{author}{\bibinfo{person}{Martin~J Beckmann},
  \bibinfo{person}{Charles~B McGuire}, {and} \bibinfo{person}{Christopher~B
  Winsten}.} \bibinfo{year}{1955}\natexlab{}.
\newblock \showarticletitle{Studies in the Economics of Transportation}.
\newblock  (\bibinfo{year}{1955}).
\newblock


\bibitem[\protect\citeauthoryear{Ben-Akiva and Lerman}{Ben-Akiva and
  Lerman}{2018}]%
        {ben2018discrete}
\bibfield{author}{\bibinfo{person}{Moshe Ben-Akiva} {and}
  \bibinfo{person}{Steven~R Lerman}.} \bibinfo{year}{2018}\natexlab{}.
\newblock \bibinfo{booktitle}{\emph{Discrete choice analysis: theory and
  application to travel demand}}.
\newblock \bibinfo{publisher}{Transportation Studies}.
\newblock


\bibitem[\protect\citeauthoryear{Biyik, Huynh, Kochenderfer, and Sadigh}{Biyik
  et~al\mbox{.}}{2020}]%
        {biyik2020active}
\bibfield{author}{\bibinfo{person}{Erdem Biyik}, \bibinfo{person}{Nicolas
  Huynh}, \bibinfo{person}{Mykel~J. Kochenderfer}, {and} \bibinfo{person}{Dorsa
  Sadigh}.} \bibinfo{year}{2020}\natexlab{}.
\newblock \showarticletitle{Active Preference-Based Gaussian Process Regression
  for Reward Learning}. In \bibinfo{booktitle}{\emph{Proceedings of Robotics:
  Science and Systems (RSS)}}.
\newblock


\bibitem[\protect\citeauthoryear{B{\i}y{\i}k, Lazar, Pedarsani, and
  Sadigh}{B{\i}y{\i}k et~al\mbox{.}}{2018}]%
        {biyik2018altruistic}
\bibfield{author}{\bibinfo{person}{Erdem B{\i}y{\i}k},
  \bibinfo{person}{Daniel~A Lazar}, \bibinfo{person}{Ramtin Pedarsani}, {and}
  \bibinfo{person}{Dorsa Sadigh}.} \bibinfo{year}{2018}\natexlab{}.
\newblock \showarticletitle{Altruistic Autonomy: Beating Congestion on Shared
  Roads}. In \bibinfo{booktitle}{\emph{Workshop on Algorithmic Foundations of
  Robotics (WAFR)}}.
\newblock


\bibitem[\protect\citeauthoryear{B{\i}y{\i}k, Lazar, Sadigh, and
  Pedarsani}{B{\i}y{\i}k et~al\mbox{.}}{2019}]%
        {biyik2019green}
\bibfield{author}{\bibinfo{person}{Erdem B{\i}y{\i}k},
  \bibinfo{person}{Daniel~A. Lazar}, \bibinfo{person}{Dorsa Sadigh}, {and}
  \bibinfo{person}{Ramtin Pedarsani}.} \bibinfo{year}{2019}\natexlab{}.
\newblock \showarticletitle{The Green Choice: Learning and Influencing Human
  Decisions on Shared Roads}. In \bibinfo{booktitle}{\emph{Proceedings of the
  58th IEEE Conference on Decision and Control (CDC)}}.
\newblock


\bibitem[\protect\citeauthoryear{B{\i}y{\i}k, Palan, Landolfi, Losey, and
  Sadigh}{B{\i}y{\i}k et~al\mbox{.}}{2020}]%
        {biyik2019asking}
\bibfield{author}{\bibinfo{person}{Erdem B{\i}y{\i}k},
  \bibinfo{person}{Malayandi Palan}, \bibinfo{person}{Nicholas~C. Landolfi},
  \bibinfo{person}{Dylan~P. Losey}, {and} \bibinfo{person}{Dorsa Sadigh}.}
  \bibinfo{year}{2020}\natexlab{}.
\newblock \showarticletitle{Asking Easy Questions: A User-Friendly Approach to
  Active Reward Learning} \emph{(\bibinfo{series}{Proceedings of Machine
  Learning Research}, Vol.~\bibinfo{volume}{100})},
  \bibfield{editor}{\bibinfo{person}{Leslie~Pack Kaelbling},
  \bibinfo{person}{Danica Kragic}, {and} \bibinfo{person}{Komei Sugiura}}
  (Eds.). \bibinfo{publisher}{PMLR}, \bibinfo{pages}{1177--1190}.
\newblock
\urldef\tempurl%
\url{http://proceedings.mlr.press/v100/b-iy-ik20a.html}
\showURL{%
\tempurl}


\bibitem[\protect\citeauthoryear{Brown and Marden}{Brown and Marden}{2016}]%
        {brown2016robustness}
\bibfield{author}{\bibinfo{person}{Philip~N Brown} {and}
  \bibinfo{person}{Jason~R Marden}.} \bibinfo{year}{2016}\natexlab{}.
\newblock \showarticletitle{The robustness of marginal-cost taxes in affine
  congestion games}.
\newblock \bibinfo{journal}{\emph{IEEE Trans. Automat. Control}}
  \bibinfo{volume}{62}, \bibinfo{number}{8} (\bibinfo{year}{2016}),
  \bibinfo{pages}{3999--4004}.
\newblock


\bibitem[\protect\citeauthoryear{Carpenter}{Carpenter}{2020}]%
        {carpenter2020la}
\bibfield{author}{\bibinfo{person}{Susan Carpenter}.}
  \bibinfo{year}{2020}\natexlab{}.
\newblock \showarticletitle{LA Traffic Could Be Worse Post Pandemic}.
\newblock \bibinfo{journal}{\emph{Spectrum News 1}} (\bibinfo{date}{Jul}
  \bibinfo{year}{2020}).
\newblock
\urldef\tempurl%
\url{https://spectrumnews1.com/ca/la-west/transportation/2020/07/30/la-traffic-could-be-worse-post-pandemic}
\showURL{%
\tempurl}


\bibitem[\protect\citeauthoryear{Correa, Schulz, and Stier-Moses}{Correa
  et~al\mbox{.}}{2008}]%
        {correa2008geometric}
\bibfield{author}{\bibinfo{person}{Jos{\'e}~R Correa},
  \bibinfo{person}{Andreas~S Schulz}, {and} \bibinfo{person}{Nicol{\'a}s~E
  Stier-Moses}.} \bibinfo{year}{2008}\natexlab{}.
\newblock \showarticletitle{A geometric approach to the price of anarchy in
  nonatomic congestion games}.
\newblock \bibinfo{journal}{\emph{Games and Economic Behavior}}
  \bibinfo{volume}{64}, \bibinfo{number}{2} (\bibinfo{year}{2008}),
  \bibinfo{pages}{457--469}.
\newblock


\bibitem[\protect\citeauthoryear{Cover}{Cover}{1999}]%
        {cover1999elements}
\bibfield{author}{\bibinfo{person}{Thomas~M Cover}.}
  \bibinfo{year}{1999}\natexlab{}.
\newblock \bibinfo{booktitle}{\emph{Elements of information theory}}.
\newblock \bibinfo{publisher}{John Wiley \& Sons}.
\newblock


\bibitem[\protect\citeauthoryear{Cui, Zhu, Wang, Wang, Zhou, Cao, Kopca, and
  Wang}{Cui et~al\mbox{.}}{2020}]%
        {cui2020traffic}
\bibfield{author}{\bibinfo{person}{Zhiyong Cui}, \bibinfo{person}{Meixin Zhu},
  \bibinfo{person}{Shuo Wang}, \bibinfo{person}{Pengfei Wang},
  \bibinfo{person}{Yang Zhou}, \bibinfo{person}{Qianxia Cao},
  \bibinfo{person}{Cole Kopca}, {and} \bibinfo{person}{Yinhai Wang}.}
  \bibinfo{year}{2020}\natexlab{}.
\newblock \showarticletitle{Traffic Performance Score for Measuring the Impact
  of COVID-19 on Urban Mobility}.
\newblock \bibinfo{journal}{\emph{arXiv preprint arXiv:2007.00648}}
  (\bibinfo{year}{2020}).
\newblock


\bibitem[\protect\citeauthoryear{Daskin}{Daskin}{1985}]%
        {daskin1985urban}
\bibfield{author}{\bibinfo{person}{Mark~S Daskin}.}
  \bibinfo{year}{1985}\natexlab{}.
\newblock \bibinfo{title}{Urban transportation networks: Equilibrium analysis
  with mathematical programming methods}.
\newblock
\newblock


\bibitem[\protect\citeauthoryear{Dowling, Singh, and Wei-Kuo~Cheng}{Dowling
  et~al\mbox{.}}{1998}]%
        {dowling1998accuracy}
\bibfield{author}{\bibinfo{person}{Richard~G Dowling},
  \bibinfo{person}{Rupinder Singh}, {and} \bibinfo{person}{Willis
  Wei-Kuo~Cheng}.} \bibinfo{year}{1998}\natexlab{}.
\newblock \showarticletitle{Accuracy and performance of improved speed-flow
  curves}.
\newblock \bibinfo{journal}{\emph{Transportation research record}}
  \bibinfo{volume}{1646}, \bibinfo{number}{1} (\bibinfo{year}{1998}),
  \bibinfo{pages}{9--17}.
\newblock


\bibitem[\protect\citeauthoryear{Ewoldsen}{Ewoldsen}{2020}]%
        {ewoldsen2020covid}
\bibfield{author}{\bibinfo{person}{Beth Ewoldsen}.}
  \bibinfo{year}{2020}\natexlab{}.
\newblock \showarticletitle{COVID-19 Trends Impacting the Future of
  Transportation Planning and Research}.
\newblock \bibinfo{journal}{\emph{National Academies of Sciences, Engineering,
  and Medicine}} (\bibinfo{date}{Aug} \bibinfo{year}{2020}).
\newblock
\urldef\tempurl%
\url{https://www.nationalacademies.org/trb/blog/covid-19-trends-impacting-the-future-of-transportation-planning-and-research}
\showURL{%
\tempurl}


\bibitem[\protect\citeauthoryear{Fleischer, Jain, and Mahdian}{Fleischer
  et~al\mbox{.}}{2004}]%
        {fleischer2004tolls}
\bibfield{author}{\bibinfo{person}{Lisa Fleischer}, \bibinfo{person}{Kamal
  Jain}, {and} \bibinfo{person}{Mohammad Mahdian}.}
  \bibinfo{year}{2004}\natexlab{}.
\newblock \showarticletitle{Tolls for heterogeneous selfish users in
  multicommodity networks and generalized congestion games}. In
  \bibinfo{booktitle}{\emph{45th Annual IEEE Symposium on Foundations of
  Computer Science}}. IEEE, \bibinfo{pages}{277--285}.
\newblock


\bibitem[\protect\citeauthoryear{Florian and Nguyen}{Florian and
  Nguyen}{1976}]%
        {florian1976recent}
\bibfield{author}{\bibinfo{person}{Michael Florian} {and} \bibinfo{person}{Sang
  Nguyen}.} \bibinfo{year}{1976}\natexlab{}.
\newblock \showarticletitle{Recent experience with equilibrium methods for the
  study of a congested urban area}.
\newblock In \bibinfo{booktitle}{\emph{Traffic Equilibrium Methods}}.
  \bibinfo{publisher}{Springer}, \bibinfo{pages}{382--395}.
\newblock


\bibitem[\protect\citeauthoryear{Golovin and Krause}{Golovin and
  Krause}{2011}]%
        {golovin2011adaptive}
\bibfield{author}{\bibinfo{person}{Daniel Golovin} {and}
  \bibinfo{person}{Andreas Krause}.} \bibinfo{year}{2011}\natexlab{}.
\newblock \showarticletitle{Adaptive submodularity: Theory and applications in
  active learning and stochastic optimization}.
\newblock \bibinfo{journal}{\emph{Journal of Artificial Intelligence Research}}
   \bibinfo{volume}{42} (\bibinfo{year}{2011}), \bibinfo{pages}{427--486}.
\newblock


\bibitem[\protect\citeauthoryear{Hattrup-Silberberg, Hausler, Heineke, Laverty,
  Möller, Schwedhelm, and Wu}{Hattrup-Silberberg et~al\mbox{.}}{2020}]%
        {hattrup2020five}
\bibfield{author}{\bibinfo{person}{Martin Hattrup-Silberberg},
  \bibinfo{person}{Saskia Hausler}, \bibinfo{person}{Kersten Heineke},
  \bibinfo{person}{Nicholas Laverty}, \bibinfo{person}{Timo Möller},
  \bibinfo{person}{Dennis Schwedhelm}, {and} \bibinfo{person}{Ting Wu}.}
  \bibinfo{year}{2020}\natexlab{}.
\newblock \showarticletitle{Five COVID-19 aftershocks reshaping mobility’s
  future}.
\newblock \bibinfo{journal}{\emph{McKinsey \& Company}} (\bibinfo{date}{Sep}
  \bibinfo{year}{2020}).
\newblock
\urldef\tempurl%
\url{https://www.mckinsey.com/industries/automotive-and-assembly/our-insights/five-covid-19-aftershocks-reshaping-mobilitys-future}
\showURL{%
\tempurl}


\bibitem[\protect\citeauthoryear{Hertzke, Middleton, Neu, and Weaver}{Hertzke
  et~al\mbox{.}}{2020}]%
        {hertzke2020moving}
\bibfield{author}{\bibinfo{person}{Patrick Hertzke}, \bibinfo{person}{Simon
  Middleton}, \bibinfo{person}{Guillaume Neu}, {and} \bibinfo{person}{Henry
  Weaver}.} \bibinfo{year}{2020}\natexlab{}.
\newblock \showarticletitle{Moving forward: How COVID-19 will affect mobility
  in the United Kingdom}.
\newblock \bibinfo{journal}{\emph{McKinsey \& Company}} (\bibinfo{date}{June}
  \bibinfo{year}{2020}).
\newblock
\urldef\tempurl%
\url{https://www.mckinsey.com/industries/automotive-and-assembly/our-insights/moving-forward-how-covid-19-will-affect-mobility-in-the-united-kingdom}
\showURL{%
\tempurl}


\bibitem[\protect\citeauthoryear{Hu and Schweber}{Hu and Schweber}{2020}]%
        {hu2020will}
\bibfield{author}{\bibinfo{person}{Winnie Hu} {and} \bibinfo{person}{Nate
  Schweber}.} \bibinfo{year}{2020}\natexlab{}.
\newblock \showarticletitle{Will Cars Rule the Roads in Post-Pandemic New
  York?}
\newblock \bibinfo{journal}{\emph{The New York Times}} (\bibinfo{date}{Aug}
  \bibinfo{year}{2020}).
\newblock
\urldef\tempurl%
\url{https://www.nytimes.com/2020/08/10/nyregion/nyc-streets-parking-dining-busways.html}
\showURL{%
\tempurl}


\bibitem[\protect\citeauthoryear{Hu, Barbour, Samaranayake, and Work}{Hu
  et~al\mbox{.}}{2020a}]%
        {hu2020impacts}
\bibfield{author}{\bibinfo{person}{Yue Hu}, \bibinfo{person}{William Barbour},
  \bibinfo{person}{Samitha Samaranayake}, {and} \bibinfo{person}{Dan Work}.}
  \bibinfo{year}{2020}\natexlab{a}.
\newblock \showarticletitle{Impacts of Covid-19 mode shift on road traffic}.
\newblock \bibinfo{journal}{\emph{arXiv preprint arXiv:2005.01610}}
  (\bibinfo{year}{2020}).
\newblock


\bibitem[\protect\citeauthoryear{Hu, Barbour, Samaranayake, and Work}{Hu
  et~al\mbox{.}}{2020b}]%
        {hu2020the}
\bibfield{author}{\bibinfo{person}{Yue Hu}, \bibinfo{person}{Will Barbour},
  \bibinfo{person}{Samitha Samaranayake}, {and} \bibinfo{person}{Dan Work}.}
  \bibinfo{year}{2020}\natexlab{b}.
\newblock \showarticletitle{The rebound --- How Covid-19 could lead to worse
  traffic}.
\newblock \bibinfo{journal}{\emph{Medium}} (\bibinfo{date}{Apr}
  \bibinfo{year}{2020}).
\newblock
\urldef\tempurl%
\url{https://medium.com/@barbourww/the-rebound-how-covid-19-could-lead-to-worse-traffic-cb245a5b1da2}
\showURL{%
\tempurl}


\bibitem[\protect\citeauthoryear{Kahaner}{Kahaner}{2020}]%
        {kahaner2020working}
\bibfield{author}{\bibinfo{person}{Larry Kahaner}.}
  \bibinfo{year}{2020}\natexlab{}.
\newblock \showarticletitle{Working from home a major factor in post-pandemic
  traffic}.
\newblock \bibinfo{journal}{\emph{FleetOwner}} (\bibinfo{date}{Aug}
  \bibinfo{year}{2020}).
\newblock
\urldef\tempurl%
\url{https://www.fleetowner.com/covid-19-coverage/article/21139861/working-from-home-a-major-factor-in-postpandemic-traffic}
\showURL{%
\tempurl}


\bibitem[\protect\citeauthoryear{Kallus and Udell}{Kallus and Udell}{2016}]%
        {kallus2016revealed}
\bibfield{author}{\bibinfo{person}{Nathan Kallus} {and}
  \bibinfo{person}{Madeleine Udell}.} \bibinfo{year}{2016}\natexlab{}.
\newblock \showarticletitle{Revealed preference at scale: Learning personalized
  preferences from assortment choices}. In
  \bibinfo{booktitle}{\emph{Proceedings of the 2016 ACM Conference on Economics
  and Computation}}. \bibinfo{pages}{821--837}.
\newblock


\bibitem[\protect\citeauthoryear{Kang, Gao, Liang, Li, Rao, and Kruse}{Kang
  et~al\mbox{.}}{2020}]%
        {kang2020multiscale}
\bibfield{author}{\bibinfo{person}{Yuhao Kang}, \bibinfo{person}{Song Gao},
  \bibinfo{person}{Yunlei Liang}, \bibinfo{person}{Mingxiao Li},
  \bibinfo{person}{Jinmeng Rao}, {and} \bibinfo{person}{Jake Kruse}.}
  \bibinfo{year}{2020}\natexlab{}.
\newblock \showarticletitle{Multiscale dynamic human mobility flow dataset in
  the us during the covid-19 epidemic}.
\newblock \bibinfo{journal}{\emph{arXiv preprint arXiv:2008.12238}}
  (\bibinfo{year}{2020}).
\newblock


\bibitem[\protect\citeauthoryear{Koppelman}{Koppelman}{1983}]%
        {koppelman1983predicting}
\bibfield{author}{\bibinfo{person}{Frank~S Koppelman}.}
  \bibinfo{year}{1983}\natexlab{}.
\newblock \showarticletitle{Predicting transit ridership in response to transit
  service changes}.
\newblock \bibinfo{journal}{\emph{Journal of Transportation Engineering}}
  \bibinfo{volume}{109}, \bibinfo{number}{4} (\bibinfo{year}{1983}),
  \bibinfo{pages}{548--564}.
\newblock


\bibitem[\protect\citeauthoryear{Kraft}{Kraft}{1988}]%
        {slsqp}
\bibfield{author}{\bibinfo{person}{Dieter Kraft}.}
  \bibinfo{year}{1988}\natexlab{}.
\newblock \bibinfo{booktitle}{\emph{A Software Package for Sequential Quadratic
  Programming}}.
\newblock \bibinfo{type}{{T}echnical {R}eport}. \bibinfo{institution}{Institut
  f{\"u}r Dynamik der Flugsysteme Oberpfaffenhofen}.
\newblock


\bibitem[\protect\citeauthoryear{Krichene, Reilly, Amin, and Bayen}{Krichene
  et~al\mbox{.}}{2018}]%
        {krichene2017stackelberg}
\bibfield{author}{\bibinfo{person}{Walid Krichene}, \bibinfo{person}{Jack~D
  Reilly}, \bibinfo{person}{Saurabh Amin}, {and} \bibinfo{person}{Alexandre~M
  Bayen}.} \bibinfo{year}{2018}\natexlab{}.
\newblock \showarticletitle{Stackelberg routing on parallel transportation
  networks}.
\newblock \bibinfo{journal}{\emph{Handbook of Dynamic Game Theory}}
  (\bibinfo{year}{2018}).
\newblock


\bibitem[\protect\citeauthoryear{Kucharski and Drabicki}{Kucharski and
  Drabicki}{2017}]%
        {kucharski2017estimating}
\bibfield{author}{\bibinfo{person}{Rafa{\l} Kucharski} {and}
  \bibinfo{person}{Arkadiusz Drabicki}.} \bibinfo{year}{2017}\natexlab{}.
\newblock \showarticletitle{Estimating macroscopic volume delay functions with
  the traffic density derived from measured speeds and flows}.
\newblock \bibinfo{journal}{\emph{Journal of Advanced Transportation}}
  \bibinfo{volume}{2017} (\bibinfo{year}{2017}).
\newblock


\bibitem[\protect\citeauthoryear{Lazar, Coogan, and Pedarsani}{Lazar
  et~al\mbox{.}}{2020}]%
        {lazar2020routing}
\bibfield{author}{\bibinfo{person}{Daniel Lazar}, \bibinfo{person}{Samuel
  Coogan}, {and} \bibinfo{person}{Ramtin Pedarsani}.}
  \bibinfo{year}{2020}\natexlab{}.
\newblock \showarticletitle{Routing for traffic networks with mixed autonomy}.
\newblock \bibinfo{journal}{\emph{IEEE Trans. Automat. Control}}
  (\bibinfo{year}{2020}).
\newblock


\bibitem[\protect\citeauthoryear{Lazar, B{\i}y{\i}k, Sadigh, and
  Pedarsani}{Lazar et~al\mbox{.}}{2019a}]%
        {lazar2019learning}
\bibfield{author}{\bibinfo{person}{Daniel~A Lazar}, \bibinfo{person}{Erdem
  B{\i}y{\i}k}, \bibinfo{person}{Dorsa Sadigh}, {and} \bibinfo{person}{Ramtin
  Pedarsani}.} \bibinfo{year}{2019}\natexlab{a}.
\newblock \showarticletitle{Learning How to Dynamically Route Autonomous
  Vehicles on Shared Roads}.
\newblock \bibinfo{journal}{\emph{arXiv preprint arXiv:1909.03664}}
  (\bibinfo{year}{2019}).
\newblock


\bibitem[\protect\citeauthoryear{Lazar, Coogan, and Pedarsani}{Lazar
  et~al\mbox{.}}{2019b}]%
        {lazar2019optimal}
\bibfield{author}{\bibinfo{person}{Daniel~A Lazar}, \bibinfo{person}{Samuel
  Coogan}, {and} \bibinfo{person}{Ramtin Pedarsani}.}
  \bibinfo{year}{2019}\natexlab{b}.
\newblock \showarticletitle{Optimal tolling for heterogeneous traffic networks
  with mixed autonomy}. In \bibinfo{booktitle}{\emph{2019 IEEE 58th Conference
  on Decision and Control (CDC)}}. IEEE, \bibinfo{pages}{4103--4108}.
\newblock


\bibitem[\protect\citeauthoryear{Mehr and Horowitz}{Mehr and Horowitz}{2019}]%
        {mehr2019will}
\bibfield{author}{\bibinfo{person}{Negar Mehr} {and} \bibinfo{person}{Roberto
  Horowitz}.} \bibinfo{year}{2019}\natexlab{}.
\newblock \showarticletitle{How will the presence of autonomous vehicles affect
  the equilibrium state of traffic networks?}
\newblock \bibinfo{journal}{\emph{IEEE Transactions on Control of Network
  Systems}} \bibinfo{volume}{7}, \bibinfo{number}{1} (\bibinfo{year}{2019}),
  \bibinfo{pages}{96--105}.
\newblock


\bibitem[\protect\citeauthoryear{of~Public~Roads}{of~Public~Roads}{1964}]%
        {united1964traffic}
\bibfield{author}{\bibinfo{person}{United States.~Bureau of Public~Roads}.}
  \bibinfo{year}{1964}\natexlab{}.
\newblock \bibinfo{booktitle}{\emph{Traffic assignment manual}}.
\newblock \bibinfo{publisher}{US Department of Commerce, Bureau of Public
  Roads, Office of Planning, Urban Planning Division}.
\newblock


\bibitem[\protect\citeauthoryear{Pascale}{Pascale}{2020}]%
        {pascale2020here}
\bibfield{author}{\bibinfo{person}{Jordan Pascale}.}
  \bibinfo{year}{2020}\natexlab{}.
\newblock \showarticletitle{Here Are Four Charts That Show The Pandemic’s
  Impact On Locals’ Travel Habits}.
\newblock \bibinfo{journal}{\emph{Wamu}} (\bibinfo{date}{Jul}
  \bibinfo{year}{2020}).
\newblock
\urldef\tempurl%
\url{https://wamu.org/story/20/07/16/here-are-four-charts-that-show-the-pandemics-impact-on-locals-travel-habits/}
\showURL{%
\tempurl}


\bibitem[\protect\citeauthoryear{Pedersen}{Pedersen}{2020}]%
        {pedersen2020impacts}
\bibfield{author}{\bibinfo{person}{Neil Pedersen}.}
  \bibinfo{year}{2020}\natexlab{}.
\newblock \bibinfo{title}{Impacts of COVID-19 on Transportation and Key
  Considerations for the Future}.
\newblock
\newblock
\urldef\tempurl%
\url{http://onlinepubs.trb.org/onlinepubs/PedersenITEAnnual\%20Meeting200813.pptx}
\showURL{%
\tempurl}
\newblock
\shownote{Presentation at ITE 2020 Annual Meeting (Online).}


\bibitem[\protect\citeauthoryear{{Republic of Turkey Governorship of
  Istanbul}}{{Republic of Turkey Governorship of Istanbul}}{2020}]%
        {istanbul2020governor}
\bibfield{author}{\bibinfo{person}{{Republic of Turkey Governorship of
  Istanbul}}.} \bibinfo{year}{2020}\natexlab{}.
\newblock \showarticletitle{Governor Yerlikaya Made a Press Release Regarding
  Gradual Working Hours Practice in Istanbul}.
\newblock  (\bibinfo{date}{Sep} \bibinfo{year}{2020}).
\newblock
\urldef\tempurl%
\url{http://en.istanbul.gov.tr/governor-yerlikaya-made-a-press-release-regarding-gradual-working-hours-practice-in-istanbul}
\showURL{%
\tempurl}


\bibitem[\protect\citeauthoryear{Roughgarden and Tardos}{Roughgarden and
  Tardos}{2002}]%
        {roughgarden2002bad}
\bibfield{author}{\bibinfo{person}{Tim Roughgarden} {and}
  \bibinfo{person}{{\'E}va Tardos}.} \bibinfo{year}{2002}\natexlab{}.
\newblock \showarticletitle{How bad is selfish routing?}
\newblock \bibinfo{journal}{\emph{Journal of the ACM (JACM)}}
  \bibinfo{volume}{49}, \bibinfo{number}{2} (\bibinfo{year}{2002}),
  \bibinfo{pages}{236--259}.
\newblock


\bibitem[\protect\citeauthoryear{Sadigh, Dragan, Sastry, and Seshia}{Sadigh
  et~al\mbox{.}}{2017}]%
        {sadigh2017active}
\bibfield{author}{\bibinfo{person}{Dorsa Sadigh}, \bibinfo{person}{Anca~D.
  Dragan}, \bibinfo{person}{S.~Shankar Sastry}, {and}
  \bibinfo{person}{Sanjit~A. Seshia}.} \bibinfo{year}{2017}\natexlab{}.
\newblock \showarticletitle{Active Preference-Based Learning of Reward
  Functions}. In \bibinfo{booktitle}{\emph{Proceedings of Robotics: Science and
  Systems (RSS)}}.
\newblock


\bibitem[\protect\citeauthoryear{Sandholm}{Sandholm}{2002}]%
        {sandholm2002evolutionary}
\bibfield{author}{\bibinfo{person}{William~H Sandholm}.}
  \bibinfo{year}{2002}\natexlab{}.
\newblock \showarticletitle{Evolutionary implementation and congestion
  pricing}.
\newblock \bibinfo{journal}{\emph{The Review of Economic Studies}}
  \bibinfo{volume}{69}, \bibinfo{number}{3} (\bibinfo{year}{2002}),
  \bibinfo{pages}{667--689}.
\newblock


\bibitem[\protect\citeauthoryear{Swamy}{Swamy}{2012}]%
        {swamy2012effectiveness}
\bibfield{author}{\bibinfo{person}{Chaitanya Swamy}.}
  \bibinfo{year}{2012}\natexlab{}.
\newblock \showarticletitle{The effectiveness of Stackelberg strategies and
  tolls for network congestion games}.
\newblock \bibinfo{journal}{\emph{ACM Transactions on Algorithms (TALG)}}
  \bibinfo{volume}{8}, \bibinfo{number}{4} (\bibinfo{year}{2012}),
  \bibinfo{pages}{1--19}.
\newblock


\bibitem[\protect\citeauthoryear{Teale}{Teale}{2020}]%
        {teale2020transportation}
\bibfield{author}{\bibinfo{person}{Chris Teale}.}
  \bibinfo{year}{2020}\natexlab{}.
\newblock \showarticletitle{Transportation leaders focus on regaining trust
  before building anew}.
\newblock \bibinfo{journal}{\emph{Smart Cities Dive}} (\bibinfo{date}{May}
  \bibinfo{year}{2020}).
\newblock
\urldef\tempurl%
\url{https://www.smartcitiesdive.com/news/transportation-leaders-focus-on-regaining-trust-before-building-anew/577394}
\showURL{%
\tempurl}


\bibitem[\protect\citeauthoryear{Tirachini and Cats}{Tirachini and
  Cats}{2020}]%
        {tirachini2020covid}
\bibfield{author}{\bibinfo{person}{Alejandro Tirachini} {and}
  \bibinfo{person}{Oded Cats}.} \bibinfo{year}{2020}\natexlab{}.
\newblock \showarticletitle{COVID-19 and public transportation: Current
  assessment, prospects, and research needs}.
\newblock \bibinfo{journal}{\emph{Journal of Public Transportation}}
  \bibinfo{volume}{22}, \bibinfo{number}{1} (\bibinfo{year}{2020}),
  \bibinfo{pages}{1}.
\newblock


\bibitem[\protect\citeauthoryear{Vitale, Bowman, and Robinson}{Vitale
  et~al\mbox{.}}{2020}]%
        {vitale2020how}
\bibfield{author}{\bibinfo{person}{Joe Vitale}, \bibinfo{person}{Karen Bowman},
  {and} \bibinfo{person}{Ryan Robinson}.} \bibinfo{year}{2020}\natexlab{}.
\newblock \showarticletitle{How the pandemic is changing the future of
  automotive}.
\newblock \bibinfo{journal}{\emph{Deloitte}} (\bibinfo{date}{Jul}
  \bibinfo{year}{2020}).
\newblock
\urldef\tempurl%
\url{https://deloitte.com/us/en/insights/industry/retail-distribution/consumer-behavior-trends-state-of-the-consumer-tracker/future-of-automotive-industry-pandemic.html}
\showURL{%
\tempurl}


\bibitem[\protect\citeauthoryear{Wang, Wei, Lin, and Li}{Wang
  et~al\mbox{.}}{2020}]%
        {wang2020spatial}
\bibfield{author}{\bibinfo{person}{Songhe Wang}, \bibinfo{person}{Kangda Wei},
  \bibinfo{person}{Lei Lin}, {and} \bibinfo{person}{Weizi Li}.}
  \bibinfo{year}{2020}\natexlab{}.
\newblock \showarticletitle{Spatial-temporal Analysis of COVID-19's Impact on
  Human Mobility: the Case of the United States}.
\newblock \bibinfo{journal}{\emph{arXiv preprint arXiv:2010.03707}}
  (\bibinfo{year}{2020}).
\newblock


\bibitem[\protect\citeauthoryear{Westblade, Brar, Pinheiro, Paidoussis, Rajan,
  Martin, Goyal, Sepulveda, Zhang, George, et~al\mbox{.}}{Westblade
  et~al\mbox{.}}{2020}]%
        {westblade2020sars}
\bibfield{author}{\bibinfo{person}{Lars~F Westblade},
  \bibinfo{person}{Gagandeep Brar}, \bibinfo{person}{Laura~C Pinheiro},
  \bibinfo{person}{Demetrios Paidoussis}, \bibinfo{person}{Mangala Rajan},
  \bibinfo{person}{Peter Martin}, \bibinfo{person}{Parag Goyal},
  \bibinfo{person}{Jorge~L Sepulveda}, \bibinfo{person}{Lisa Zhang},
  \bibinfo{person}{Gary George}, {et~al\mbox{.}}}
  \bibinfo{year}{2020}\natexlab{}.
\newblock \showarticletitle{SARS-CoV-2 viral load predicts mortality in
  patients with and without cancer who are hospitalized with COVID-19}.
\newblock \bibinfo{journal}{\emph{Cancer cell}} (\bibinfo{year}{2020}).
\newblock


\bibitem[\protect\citeauthoryear{Wilde, Kulic, and Smith}{Wilde
  et~al\mbox{.}}{2020}]%
        {wilde2020active}
\bibfield{author}{\bibinfo{person}{Nils Wilde}, \bibinfo{person}{Dana Kulic},
  {and} \bibinfo{person}{Stephen~L Smith}.} \bibinfo{year}{2020}\natexlab{}.
\newblock \showarticletitle{Active Preference Learning using Maximum Regret}.
  In \bibinfo{booktitle}{\emph{Proceedings of the IEEE/RSJ International
  Conference on Intelligent Robots and Systems (IROS)}}.
\newblock


\bibitem[\protect\citeauthoryear{Wong and Wong}{Wong and Wong}{2016}]%
        {wong2016network}
\bibfield{author}{\bibinfo{person}{Wai Wong} {and} \bibinfo{person}{SC Wong}.}
  \bibinfo{year}{2016}\natexlab{}.
\newblock \showarticletitle{Network topological effects on the macroscopic
  Bureau of Public Roads function}.
\newblock \bibinfo{journal}{\emph{Transportmetrica A: Transport Science}}
  \bibinfo{volume}{12}, \bibinfo{number}{3} (\bibinfo{year}{2016}),
  \bibinfo{pages}{272--296}.
\newblock


\bibitem[\protect\citeauthoryear{Zadimoghaddam and Roth}{Zadimoghaddam and
  Roth}{2012}]%
        {zadimoghaddam2012efficiently}
\bibfield{author}{\bibinfo{person}{Morteza Zadimoghaddam} {and}
  \bibinfo{person}{Aaron Roth}.} \bibinfo{year}{2012}\natexlab{}.
\newblock \showarticletitle{Efficiently learning from revealed preference}. In
  \bibinfo{booktitle}{\emph{International Workshop on Internet and Network
  Economics}}. Springer, \bibinfo{pages}{114--127}.
\newblock


\bibitem[\protect\citeauthoryear{Zheng, Zhang, and Nie}{Zheng
  et~al\mbox{.}}{2020}]%
        {zheng2020fall}
\bibfield{author}{\bibinfo{person}{Hongyu Zheng}, \bibinfo{person}{Kenan
  Zhang}, {and} \bibinfo{person}{Marco Nie}.} \bibinfo{year}{2020}\natexlab{}.
\newblock \showarticletitle{The Fall and Rise of the Taxi Industry in the
  COVID-19 Pandemic: A Case Study}.
\newblock \bibinfo{journal}{\emph{Available at SSRN 3674241}}
  (\bibinfo{year}{2020}).
\newblock


\end{thebibliography}
